# CRITICISM OF "NECESSITY OF SIMULTANEOUS CO-EXISTENCE OF INSTANTANEOUS AND RETARDED INTERACTIONS IN CLASSICAL ELECTRODYNAMICS"[1]

J. D. JACKSON
*Department of Physics, University of California, Berkeley .*
*Berkeley, CA 94720, USA*

The demonstration that the electromagnetic fields derived from the Liénard-Wiechert potentials do not satisfy the Maxwell equations is proved to be false. Errors were made in the computation of the derivatives of retarded quantities. The subsequent inference of the *necessity* of both instantaneous and retarded electromagnetic interactions cannot be made. Different choices of gauge can lead to a variety of forms for the scalar and vector potentials, always with the same retarded *field*s. Classical electromagnetic theory is complete as usually expressed. One may choose to work in the Lorenz gauge in which all quantities are retarded.

## 1. Introduction

The authors of this paper[1] make the claim that the electric and magnetic fields derived from the Liénard-Wiechert potentials for a charged particle in arbitrary motion do not satisfy the Maxwell equations. In particular, they purport to show that the fields do not satisfy the Maxwell equation corresponding to Faraday's law. This claim is particularly puzzling since the structure of the two homogeneous Maxwell equations (Faraday's law and absence of magnetic charges) are what permit the introduction of the scalar and vector potentials in the first place. The expressions for the fields in terms of the potentials,

$$\vec{E} = -\vec{\nabla}\varphi - \frac{\partial \vec{A}}{c\,\partial t} \;;\quad \vec{B} = \vec{\nabla}\times\vec{A} \tag{1}$$

are such that the equations representing Faraday's law and the absence of magnetic charges,

$$\vec{\nabla}\times\vec{E} + \frac{\partial \vec{B}}{c\,\partial t} = 0 \;;\quad \vec{\nabla}\cdot\vec{B} = 0 \tag{2}$$

are satisfied automatically. Yet the authors begin with the Liénard-Wiechert potentials, which are derived from the retarded solutions in the Lorenz gauge with a point charge source, find the well-known expressions for the electric and magnetic fields, and the proceed to show that Faraday's law is not obeyed! Apparently the authors believe only selectively in the mathematics of differentiation. If one substitutes the definitions (1) into the first equation of (2) one finds trivially $\vec{\nabla}\times\vec{\nabla}\varphi = 0$. Faraday's law must be satisfied!

## 2. Source of the Mistake

How can it be that the authors find the correct expressions for the fields by differentiation of the potentials with attention to the retarded nature of the expressions, yet fail to find Faraday's law obeyed? The answer is a fluke. The authors make two errors in computing derivatives of the retarded expressions for the Liénard-Wiechert potentials. The first error occurs in their Eq.(11), first and third equations. They ignore a contribution to the spatial partial derivative for fixed time *t* from explicit spatial coordinate dependence (of the observation point), including only the implicit dependence through the retarded time $t_0$. The second error occurs in the first equation in the unnumbered set of equations below Eq.(15). It reads

$$\frac{\partial R}{\partial t_0} = -c \qquad (3)$$

where $\vec{R} = \vec{x} - \vec{x}_0(t_0)$, $R$ is the magnitude of $\vec{R}$, $t_0 = t - R(t_0)/c$, $\vec{x}$ is the observation point, and $\vec{x}_0(t_0)$ is the coordinate of the charged particle source. This first equation (3) is inconsistent with the correct second equation in that line. The correct result can be found as follows. Consider the time derivative of $R^2$ with respect to $t_0$:

$$\frac{\partial R^2}{\partial t_0} = 2R\frac{\partial R}{\partial t_0} = 2\sum_{j=1}^{3} R_j \frac{\partial R_j}{\partial t_0} = -2\vec{R}\cdot\vec{V} \qquad (4)$$

The final step results from using the authors' correct second equation in the line. Hence we find

$$\frac{\partial R}{\partial t_0} = -\frac{\vec{R}\cdot\vec{V}}{R} \qquad (5)$$

This must replace the incorrect equation (3).

### 3. The Correct Fields? How?

Given two mistakes, how did the authors managed to obtain the correct Liénard-Wiechert fields? The answer is a lucky cancellation of an error in the gradient of the scalar potential with a compensating error in the time derivative of the vector potential. Explicitly (with unit charge for simplicity), one finds a term in the gradient of the scalar potential beyond the authors:

$$\vec{\nabla}\varphi = \frac{\vec{\beta}}{(R - \vec{\beta}\cdot\vec{R})^2} + \vec{\nabla}t_0 \frac{\partial \varphi}{\partial t_0} \qquad (6)$$

The derivative with respect to $t_0$ is the authors' Eq.(16). A compensating term occurs in the time derivative of the vector potential,

$$\frac{\partial \vec{A}}{c\partial t} = \frac{-\vec{\beta}}{(R - \vec{\beta}\cdot\vec{R})^2} + \frac{\partial t_0}{\partial t}\frac{\partial \vec{A}}{c\partial t_0} \qquad (7)$$

where the derivative with respect to $t_0$ here is the authors' Eq.(18). In the sum the two added terms cancel to give them the well-known correct Liénard-Wiechert electric field.

The correct Liénard-Wiechert magnetic field emerges similarly. The correct computation of the curl of $\vec{A}$ yields

$$\vec{\nabla} \times \vec{A} = \frac{\vec{\beta} \times \vec{\dot\beta}}{(R - \vec{\beta}\cdot\vec{R})^2} + \vec{\nabla} t_0 \times \frac{\partial \vec{A}}{\partial t_0} = \vec{\nabla} t_0 \times \frac{\partial \vec{A}}{\partial t_0}. \tag{8}$$

The last form is the authors' Eq.(13).

### 4. Faraday's Law is Obeyed

The fact that the authors' errors still gave the correct results for the fields is unfortunate. Armed with a mistaken faith in their (erroneous) method of calculation, they proceed to compute the necessary derivatives to test Faraday's law. They do not show any of the intermediate steps, but they do stress that they use the incorrect (3) instead of (5). Unfortunately for them, a fortuitous cancellation does not occur. Consequently they "prove" that Faraday's law is not valid.

The Liénard-Wiechert fields are

$$E_k = \frac{1}{(\kappa R)^3}(1 - \beta^2 + \vec{R}\cdot\vec{\dot\beta}/c)(R_k - R\beta_k) - (\kappa R) R \dot\beta_k/c \tag{9}$$

$$B_i = \varepsilon_{ijk} R_j E_k / R \tag{10}$$

where $\kappa = 1 - \vec{R}\cdot\vec{\beta}/R$, $\vec{\beta} = d\vec{x}_0(t_0)/cdt_0$, $\vec{\dot\beta} = d\vec{\beta}/dt_0$ are convenient shorthands, and all quantities are to be evaluated at the retarded time $t_0 = t - R(t_0)/c$. Unlike the potentials they are not merely functions of $(R - \vec{\beta}\cdot\vec{R})$ and $\vec{\beta}$. Other derivatives enter. While the result of the substitution of the fields (9) and (10) into the equations (2) is a foregone conclusion (if one believes in mathematics), the explicit verification needs the following full set of derivatives:

$$\frac{\partial R}{\partial x_j} = \frac{R_j}{\kappa R}\ ; \quad \frac{\partial R_k}{\partial x_j} = \delta_{jk} + \frac{\beta_k R_j}{\kappa R}\ ; \quad \frac{\partial \beta_k}{\partial x_j} = -\frac{R_j \dot\beta_k}{c\kappa R}$$

$$\tag{11}$$

$$\frac{\partial (\kappa R)}{\partial x_j} = \frac{R_j}{\kappa R}\left[1 - \beta^2 + \vec{R}\cdot\dot{\vec{\beta}}/c\right] - \beta_j$$

and

$$\frac{\partial R}{c\partial t} = -\frac{\vec{\beta}\cdot\vec{R}}{\kappa R} \quad ; \quad \frac{\partial R_k}{c\partial t} = -\beta_k/\kappa \quad ; \quad \frac{\partial \beta_k}{c\partial t} = \dot{\beta}_k/c\kappa$$

(12)

$$\frac{\partial (\kappa R)}{c\partial t} = -\frac{\vec{\beta}\cdot\vec{R}}{\kappa R} - \frac{\dot{\vec{\beta}}\cdot\vec{R}}{c\kappa} + \frac{\beta^2}{\kappa}$$

The derivatives of $\dot{\vec{\beta}}$ follow similarly.

With these derivatives some elementary but tedious algebra shows that the fields (9) and (10) do indeed satisfy both of the homogeneous Maxwell equations (2), as well as the other two Maxwell equations.

## 5. Failure of Subsequent Arguments

The authors argue that the failure of the Liénard-Wiechert fields to satisfy the Maxwell equations implies the necessity for modification of electromagnetic theory along the lines of an earlier paper.[2] In that paper one of the authors and another collaborator consider a charge in motion in a straight line with constant acceleration. They purport to show that the parallel component of the Liénard-Wiechert electric field (which they write down correctly) does not satisfy the wave equation on the line of motion. Now, this is again evidence of a selective belief in mathematics. The Liénard-Wiechert fields are obtained by differentiation from potentials that are retarded solutions of the wave equation. Starting with the wave equations for the scalar and vector potentials, one can by suitable differentiation and allowable interchange of orders of differentiation reach the conclusion that the electromagnetic fields derived from those potentials do (must!) satisfy the wave equation. Double differentiation of the fields (9) and (10) leads to the same result. All components of the fields satisfy the wave equation.

In arriving at their false conclusion, these earlier authors[2] do not show any intermediate steps, but they do write down at least one incorrect differentiation chain rule. On the basis of their wrong result, they argue for the *necessity* of both conventional retarded interactions and instantaneous, action-at-a distance interactions. Since the original demonstration is incorrect, the subsequent arguments have no basis. It is of course known that in certain gauges the potentials can contain both retarded and instantaneous contributions[a]. But there is *no necessity* for such a mixture. And the fields are always retarded, except in the strict static limit. Electromagnetic theory is complete in any chosen gauge. In the familiar Lorenz gauge all quantities, potentials and fields, are retarded.

___________________________________

[a] A paper addressing this issue and the related misconception that the *fields* obtained from potentials of different gauges can be different is to appear elsewhere.

___________________________________

**References**
1. A. E. Chubykalo and S. J. Vlaev, *Int. J. Mod. Phys.* **A14**, 3789 (1999).
2. A. E. Chubykalo and R. Smirov-Rueda, *Phys. Rev.* **E53**, 5373 (1996)

**Proposed Running head**
    Criticism of Chubykalo and Vlaev's papers